\documentclass[a4paper,12pt]{article}
\newcommand{\beqn}{\begin{equation}}
\newcommand{\eeqn}{\end{equation}}
\newcommand{\barr}[1]{\begin{array}{#1}}
\newcommand{\earr}{\end{array}}
\newcommand{\beqna}{\begin{eqnarray}}
\newcommand{\eeqna}{\end{eqnarray}}
\newcommand{\lapprox}{\stackrel{<}{\scriptstyle \sim}}
\newcommand{\rapprox}{\stackrel{>}{\scriptstyle \sim}}
\def\question#1{}

\begin{document}
\title{
\begin{flushright}
\small{hep-ph/0309253} 
\\ \small{LA-UR-03-7076}
\end{flushright}
\vspace{0.6cm}
\Large\bf The $D^{\ast0}\bar{D}^0$ threshold resonance}
\vskip 0.2 in
\author{Frank E. Close\thanks{\small \em E-mail: F.Close1@physics.ox.ac.uk} \\
{\small \em Department of Physics - Theoretical Physics,}\\
{\small \em University of 
Oxford, 1 Keble Road, Oxford OX1 3NP, U.K. }\vspace{.3cm} \\ 
Philip R. Page\thanks{\small \em E-mail: prp@lanl.gov} \\
{\small \em Theoretical Division, MS B283, Los Alamos
National Laboratory,}\\
{\small \em  Los Alamos, NM 87545, U.S.A.}}
\date{}
\maketitle
\begin{abstract}{
Tests are discussed to
distinguish $c\bar{c}$, hybrid charmonium and molecular 
interpretations of the narrow
Belle resonance at $3872$ MeV.
}
\end{abstract}
\bigskip

PACS number(s): \hspace{.2cm} 11.30.Hv \hspace{.2cm} 12.39.Mk 
\hspace{.2cm} 13.25.Gv \hspace{.2cm} 14.40.Gx


\vspace{1.5cm}

The Belle Collaboration recently reported the $10.3$ $\sigma$ discovery
of a resonance at mass $3872.0 \pm 0.6 \pm 0.5$ MeV with a width less
than $2.3$ MeV in $J/\psi \; \pi^+ \pi^-$~\cite{belle}. The 
resonance, which is
denoted here as $X(3872)$, is produced via the decay $B^{\pm}
\to K^\pm X(3872)$~\cite{belle}.

The most remarkable feature of $X(3872)$ is that it is, within errors,
exactly at the $D^{\ast 0} \bar{D}^0$ threshold at $3871.5 \pm 0.5$
MeV~\cite{pdgnew}. In fact, $M(X) - M({D^{\ast 0}
\bar{D}^0}) = 0.5 \pm 0.9$ MeV.  The next nearest open charm
thresholds are $D^{\pm\ast} D^\mp$, which is $8.0\pm 1.0$ MeV
above $D^{\ast 0} \bar{D}^0$, and $D_s^{\pm} D_s^\mp$, $64.7\pm 1.0$
MeV above $D^{\ast 0} \bar{D}^0$~\cite{pdgnew}.  Based on the mass of
$X(3872)$ alone, it is expected that the resonance has a much larger
$D^{\ast 0} \bar{D}^0$ component in its wave function than
$D^{\pm\ast} D^\mp$, or other, components.  Even if $X(3872)$ is
hypothesized to be a $c\bar{c}$ state, the degeneracy with the ${D^{\ast
0} \bar{D}^0}$ threshold leads one to expect that the resonance couples, and
mixes, with $u\bar{u}$ more strongly than with $d\bar{d}$
since the $D^{\ast 0}$ and $D^{0}$ have
quark structure $c\bar{u}$. Hence the multiquark quark content of the state is
dominantly

\beqn
\label{iso} c\bar{c}u\bar{u} = \frac{1}{\sqrt{2}}\;
c\bar{c}\left( \frac{u\bar{u}+d\bar{d}}{\sqrt{2}} +
\frac{u\bar{u}-d\bar{d}}{\sqrt{2}}\right) = \frac{1}{\sqrt{2}} (
|I_s=0\rangle+|I_s=1\rangle ) \; ,
\eeqn
which means that the state breaks isospin symmetry maximally.  
This could turn out to
be the largest isospin breaking in the hadronic
spectrum to date. Eq.~\ref{iso} implies that the resonance has no
definite isospin, and hence no well-defined G-parity.  Isospin
symmetry has also been hypothesized to be broken via a similar 
mechanism
for the $f_0(980)$ and $a_0(980)$ states~\cite{f0mix,ct02} and for the 
$D_s(2.32;2.46)$~\cite{bcl03}.

The observed decay $X(3872)\to J/\psi \: \pi^+ \pi^-$ is not very
restrictive for the possible quantum numbers of $X$: It is only
possible to show that $X$ cannot be $J^{PC}=0^{--}$ exotic by 
conservation
of these quantum numbers in QCD.

There are preliminary indications that $X(3872)$ prefers to decay
to the high-mass part of the $\pi\pi$ spectrum in 
$J/\psi\: \pi^+ \pi^-$~\cite{belle}.
Assuming this is not due to the 
 Adler zero which is known to suppress the low-mass $\pi\pi$ spectrum 
in $\psi' \to J / \psi\: \pi \pi$, this could be evidence for the decay
$J/\psi\: \rho^0$. 
(The $J/\psi\: \omega$ threshold is $8$ MeV above $X$, so that this mode is
negligible).  Decay to $J/\psi\: \rho^0$ means that $X$
decays through its
isospin $1$ component, and has $C$-parity positive. The $J/\psi\: \rho^0$
threshold is only $6.4\pm 1.1$ MeV below the mass of the 
$X$~\cite{pdgnew}, so that
$X\to J/\psi\: \rho^0$ should preferably occur in S-wave.
If $X$ decays to $J/\psi\: \rho^0$ it cannot decay to $J/\psi\; (\pi
\pi)_S$, since this final state has negative $C$-parity.
The experimental data are consistent with $X$ not decaying to
$J/\psi\; (\pi^+ \pi^-)_S$~\cite{belle}. If $X$ indeed
decays to $J/\psi\: \rho^0$, and it is assumed that it is narrow because
it couples weakly to the only kinematically allowed open charm
threshold ($D \bar{D}$), it follows that either

\begin{enumerate}

\item the resonance has unnatural parity $0^-, 1^+, 2^-, 3^+, \ldots$,
which cannot couple to $D\bar{D}$ by conservation of $J^P$. Together
with positive $C$-parity this gives its $J^{PC} = 0^{-+}, 1^{++}, 
2^{-+},
3^{++}, \ldots$ Only $1^{++}$ can decay to $J/\psi\: \rho^0$ in S-wave.

\item the resonance is in the $J^{PC}$ exotic sequence $0^{+-},
1^{-+}, 2^{+-}, 3^{-+},\ldots$ which cannot decay to $D\bar{D}$ by
conservation of $CP$. Together with positive $C$-parity $X$ should be
$1^{-+},3^{-+},\ldots$ Such states cannot decay to $J/\psi\: \rho^0$ in 
S-wave.

\item the resonance decays to $D \bar{D}$, which is $\sim 138$ MeV
below the $X$, in a very high wave.
Resonances in the sequence $J^P = 3^-, 4^+, \ldots$ can decay to
$D \bar{D}$ in $F$-wave and higher. Incorporating positive $C$-parity
$J^{PC} = 3^{-+}, 4^{++}, \ldots$. These states cannot
decay to $J/\psi\: \rho^0$ in S-wave.

\item the decay of the resonance to $D\bar{D}$ is suppressed
dynamically.  An example of such a selection rule is that charmonium
hybrid meson decay to $D\bar{D}$ is exactly zero in non-relativistic
models with spin $1$ pair creation~\cite{page}. Also, a large
$D^\ast\bar{D}$ molecule will have suppressed decays to $D\bar{D}$,
because the decay is proportional to the wave function at the origin
$|\psi(0)|^2, |\psi'(0)|^2,\ldots$ in a non-relativistic formalism
appropriate for large molecules.

\end{enumerate}

 The detection of 
$X(3872)$ in $J/\psi \: \pi^+ \pi^-$ indicates that the state
contains $c\bar{c}$ pairs.  Various possibilities for the
interpretation of the state arise, keeping in mind that na\"{\i}ve
expectations will be skewed by the mass coincidence with the
$D^\ast\bar{D}$ threshold.  In particular, as discussed above,
the ${D^{\ast 0} \bar{D}^0}+$c.c component will contain both isospins
even though the state may have ``originated'' as isospin 0 conventional
or hybrid charmonium. 
 The possibilities are now listed starting
with the more conservative ones.  These possibilities can be
distinguished experimentally by measuring the $J^{PC}$ of the state.
 
{\it Conventional charmonium:} There are $3S$, $2P$,
$1D$ and $1F$
charmonia predicted in the relevant mass region, of which $2^{--}$ can
be narrow, if, as is expected, it is 
below the $DD^*$ threshold. However, the $2^{--}$ possibility may already 
be excluded by potential models~\cite{belle}. 
Within the realm of $C=+$ 
it is immediate from (1) and (3) that $3S$ charmonia are
probably $0^{-+}$, $2P$
charmonia are likely to
be $1^{++}$, that $1D$ charmonia should be $2^{-+}$, and that
$1F$ charmonia are probably $3^{++}$ or $4^{++}$.  The 3S and 
1F levels are
predicted to be at $\sim 4.1$ GeV, which is higher than the 2P and 1D levels,
and less likely to explain the mass of $X$. 

 Although the 2P $2^{++}$ state does couple to $D\bar{D}$, it does so
in D-wave, and an estimate suggests that the open charm width below
$D^\ast \bar{D}$ threshold for this state is $0-4$ MeV~\cite{page95}. 
Such a state is consistent
with the measured width of $X$, and can decay to $J/\psi\; \rho^0$
in S-wave. 

 {\it Hybrid charmonium:} The $X$ mass region is somewhat lower
than the region around $4.3$ GeV where the lightest hybrid charmonia
are located according to lattice QCD and models. The lightest hybrid
charmonia in lattice calculations are the TE hybrids with 
$J^{PC}=(0,1,2)^{-+}$ and $1^{--}$. The $0^{-+}$ and $2^{-+}$ do not couple
to $D\bar{D}$ from (1), the $1^{-+}$ not due to (2), and $1^{--}$ has
a suppressed coupling to $D\bar{D}$ from (4).

The $X$ may be a conventional or hybrid charmonium state that strongly
couples to the $D^\ast \bar{D}$ threshold, shifting it
to the threshold, where it acquires molecular character. In this
case no isospin partner of the $X$ is expected.

 {\it $D^\ast \bar{D}$ molecule:} Due to the nearness of the
resonance to the $D^\ast \bar{D}$ threshold, this is a natural
interpretation.  A $D^\ast\bar{D}$ molecule was previously
predicted~\cite{derujula,to,torn94}. 
If the resonance is below $D^\ast \bar{D}$ threshold,
it would be natural to assume that it has the $D^\ast$ and $\bar{D}$
in relative S-wave, since there is no evidence for other molecular
states nearby in mass.  Such a state can be $1^{+-}$ or $1^{++}$,
although the latter possibility is preferred by (1). Note that
the recently discovered $D_s(2460)$ is probably also $1^{++}$ and
may be similar to the $X$.
Because $M(X) - M({D^{\ast 0} \bar{D}^0}) = 0.5 \pm 0.9$ MeV, the
binding should be $\lapprox 0.4$ MeV, so that

\beqn\label{rms}
r_{\mbox{r.m.s.}} \rapprox \frac{1}{\sqrt{2\mu
E_{\mbox{binding}}}} = 7 \mbox{ fm}\; ,
\eeqn 
larger than the size Eq.~\ref{rms} gives for the deuteron 
($4$ fm for the deuteron binding energy of 2.22 MeV).
Here $\mu$ is the reduced mass of
$D^{\ast 0}$ and $\bar{D}^0$.  Because the constituents in the
molecule are separated by nuclear distances, two implications obtain:
(1) The binding is likely to be strongly influenced by
 long-distance $\pi^0$ exchange,
which is known to be attractive~\cite{torn94}, and (2) The
constituents move non-relativistically with momentum $p \lapprox
1/r_{\mbox{r.m.s.}} = 30$ MeV.  Because of the deuteron-like
character of this loosely bound two-meson molecule, 
the term ``deuson'' was suggested to
discriminate such states from molecules in atomic physics~\cite{to}.
$t$-channel $\pi^0$ exchange can happen
via $\bar{D}^0 \to \bar{D}^{\ast 0} \pi^0$ and $D^{\ast 0} \pi^0 \to
D$. Interestingly, $\pi$ exchange will not happen for a $D \bar{D}$
bound state, since the $\pi D \bar{D}$ vertex is zero by parity
conservation. This explains why $1^{++}\;\bar{D} D^{\ast}$ molecules can 
exist
without the existence of $0^+\;D \bar{D}$ molecules.


Tornqvist has argued~\cite{to} that in the positive charge conjugation
$I_s=0$ there is a strong attraction arising from the spin-isospin
factor associated with $\pi$ exchange, giving a ``relative binding
number" (RBN~\cite{to}) of $-{3}/{2}$ (attraction, $I_s=0$) and
$+{1}/{2}$ (repulsion, $I_s=1$.) Thus there is one $1^{++}$ bound
state in this limit. To see what happens as $m_d \gg m_u$ it is
instructive first to see how the RBN arise by enumerating the
individual contributions of the various $\pi$ charge states.  The
particles with their quark contents are $D^+\; (c\bar{d})$, $D^0\; (-
c \bar{u})$, $\bar{D}^0\; (u \bar{c})$, $D^-\; (d\bar{c})$, $\pi^+ \;
(u \bar{d})$ and $\pi^-\; (- d \bar{u})$. (We use $D$ to represent $D$
or $D^*$). The $\pi^0$ is $(u\bar{u} - d\bar{d})/\sqrt{2}$ in the
isospin limit, and $u\bar{u}$ when $m_d\to\infty$.  There are four
contributions in a specific time ordering, i.e. $D^0 \bar{D}^0 \to D^0
\bar{D}^0$ (with t-channel $\pi^0$ exchange through its $u\bar{u}$
component), $D^0 \bar{D}^0 \to D^+ D^-$ ($\pi^-$ exchange), $D^0
\bar{D}^0 \to D^+ D^-$ ($\pi^+$ exchange) and $D^+ D^-\to D^+ D^-$
($\pi^0$ exchange through its $d\bar{d}$ component).  By inserting the
quark contents, the amplitudes in the isospin limit are proportional
to $-1/{2},\; 1,\; 1$ and $- 1/{2}$ for the four contributions
respectively. In the limit $m_d\to\infty$ they behave as $-1,\; 1,\;
1$ and $0$ respectively.  In the isospin limit

\beqn\label{isop}
| I_s=0 \rangle  = \frac{D^0 \bar{D}^0  - D^+ D^-}{\sqrt{2}} \hspace{1cm} 
| I_s=1 \rangle  = \frac{D^0 \bar{D}^0  + D^+ D^-}{\sqrt{2}} 
\eeqn
 the
amplitude for the states in Eq.~\ref{isop} become proportional to
$(- 1/2 - 1 - 1 - 1/2 )/2  = - 3/2\; (I_s=0$ state) and
$(-1/2 + 1 + 1 - 1/2 )/2  = + 1/2\; (I_s=1$ state), as expected.
 When $m_d\to\infty$ the isospin basis is broken leaving two states,
an infinitely heavy $D^+D^-$ and a light $D^0 \bar{D}^0$. 
The exchange amplitudes are then driven by the $u\bar{u}$ exchange only.
The $D^+ D^-$ state experiences 
no splitting (fourth contribution).
In the same normalisation as above, the state $D^0 \bar{D}^0$ 
has an amplitude of $-1$ (first contribution).

Thus in this extreme there is a weakened binding at the $D^{\ast 0}
\bar{D}^0$ relative to the isospin limit and no effect at the charged
threshold.  An intermediate scenario where $m_u < m_d < \infty$ should
give repulsion of the one level and attraction of the other. In
general there is only one attractive state. This starts out as $I_s=0$
in the isospin limit and goes over into the $D^{\ast 0} \bar{D}^0$ in
the $m_d \to \infty$ limit. The conclusion is that there is only one
molecular state bound by the pion associated with the
$D^{\ast}\bar{D}$ threshold.

If the resonance $X$
is above the $D^{\ast 0} \bar{D}^0$ threshold, the $D^{\ast 0}$ and 
$\bar{D}^0$
are expected to be in a relative $L$-wave, with $L>1$, since this
will lead to an angular momentum barrier  suppressing the 
constituents from annihilating, as the decay will at
least be proportional to $|\psi'(0)|^2$. In addition, the potential must
have a form which enables the wave function to be localized, so that it
does not ``fall-apart'' to $D^{\ast 0}$ and $\bar{D}^0$.

If $X$ is indeed a molecule, 
its $D^{\ast 0}$ component should
decay with a width equal to that of $D^{\ast 0}$ (known to be $< 2.1$
MeV~\cite{pdgnew}, and likely smaller than the width of the
$D^{\ast +}$, which is $96\pm 4\pm 22$ keV~\cite{pdgnew}). 
This is consistent with the experimental bounds on
the width of the state. Also, these decay modes of the state should
derive from the decay modes of the $D^{\ast 0}$, i.e. the state should be
seen in $\bar{D}^0 (D^0 \pi^0)$ and $\bar{D}^0 (D^0 \gamma)$, and
charge conjugates. It is hence predicted that when these modes are
studied a signal will be seen at Belle, BABAR and 
CLEO.
 The relative strength of the $\bar{D}^0 (D^0 \pi^0)$ and
$\bar{D}^0 (D^0 \gamma)$ modes should be similar to the relative 
branching ratios of the $D^{\ast 0}$, i.e. 
$(61.9\pm 2.9\: \%)\; / \; (38.1\pm 2.9\: \%)$~\cite{pdgnew},
because the $D^{\ast 0}$ in the molecule is almost on-shell. 

In addition to the decay modes of the state mentioned above, there
will be dissociation modes 
where the $D^{\ast 0}$ and $\bar{D}^0$ come together at the origin,
rearranging the quarks to $c\bar{c}$ and $u\bar{u}$ pairs which
evolve to a charmonium and light meson. (Modes involving
a $c\bar{c}$ and two light quark pairs 
should be suppressed since an extra pair
creation is required, and are not considered further here. 
Also, the radiative decay mode $c\bar{c}\gamma$ is not expected
to be competitive as it requires not only a rearrangement of the
molecule to $c\bar{c}u\bar{u}$, but also electromagnetic suppression.
This is consistent with the non-observation of $X\to\chi_{c1}\gamma$ by 
Belle~\cite{belle}. 
Further, note that this mode will be forbidden if $C(X)=+$,
as advocated here).
The dissociation decay widths will be
proportional to $|\psi(0)|^2$ for an S-wave molecule, and $|\psi'(0)|^2$ for
a P-wave molecule. For light mesons such calculations in the case of
the S-wave molecules ($f_0(980)$ and $a_0(980)$)
can generate widths of order $100$  
MeV~\cite{zhang}, while for P-wave molecules the widths are
smaller~\cite{zhang}. In the likely scenario where the
state is an S-wave $1^{++}$ molecule, these modes will dominate those
mentioned in the previous paragraph. 
The modes allowed by phase space
for a such a molecule
are $\eta_c(\pi\pi)_S, J/\psi\: \rho^0, \chi_{c0}\pi^0, 
\chi_{c1}\pi^0, \chi_{c1}(\pi\pi)_S,
\chi_{c2}\pi^0$ and $\chi_{c2}(\pi\pi)_S$. 


 If $X$ is molecular in origin, there will also be short 
range interactions. These interactions can 
be further t- or u-channel processes, or s-channel processes. 
The latter are particularly interesting when the $D^{\ast}\bar{D}$
threshold
lies between two resonances.
These resonances will interact
with the threshold between them.
The contribution to the
potential for $D^{\ast}$ scattering with $\bar{D}$ through an
s-channel resonance is of the form

\beqn\label{sc}
\frac{g_{R_1 D D^\ast}^2}{q^2-m_{R_1}^2} + 
\frac{g_{R_2 D D^\ast}^2}{q^2-m_{R_2}^2} \; ,
\eeqn
neglecting the effect of widths. Here
 $g_{R D D^\ast}$ is the coupling of the resonance $R$ to $D$ and
$D^\ast$, and $m_R$ is the mass of the resonance. If the $D^\ast$
scattering with $\bar{D}$
is calculated at $q^2=m_X^2$, and $m_{R_1} < m_X < m_{R_2}$,
it is possible for the two terms to approximately cancel each other.
This may well be the case for the Belle resonance, as the binding 
energy of this resonance is so small compared to other molecular
candidates like the $f_0(980)$, $a_0(980)$ and $D_s(2.32;2.46)$ whose
binding is usually explained by assuming that either $R_1$ does not
exist, or that it couples weakly.
For example, if $X$ is $1^{++}$, the first resonance would be the
1P charmonium and the second one the 2P charmonium. 

As is evident for the discussion of the molecular origin of $X$, it
cannot be viewed in isolation: since interactions with charmonium
states occur, that implies that the effect of $D^\ast \bar{D}$ on
charmonium states should also be considered. Specializing to the case
of two charmonium resonances, $R_1$ and $R_2$, with $m_{R_1} <
m_{D^\ast \bar{D}} < m_{R_2}$, this mixing is expected to shift the
$R_2$ and $R_1$ masses. The shift in the $D^\ast \bar{D}$ threshold
can be analysed with the dynamics outlined around 
Eq.~\ref{sc}.  The charmonium states will acquire $D^\ast \bar{D}$
components. If $X$ is $1^{++}$, then $R_1$ is the
$\chi_{c1}(3510)$. Mixing with the $D^\ast \bar{D}$ threshold will
induce a $c\bar{c} n\bar{n}$ ($n\bar{n} =
(u\bar{u}+d\bar{d})/\sqrt{2}$) component in the $\chi_{c1}$ wave
function within isospin symmetry. Since the $D^{\ast 0}\bar{D}^0$
threshold is nearer to the $\chi_{c1}$ mass than the $D^{\ast
\pm}{D}^\mp$ threshold, the $c\bar{c} u\bar{u}$ component will
dominate the $c\bar{c} d\bar{d}$ component, leading to isospin
violating decays like $\chi_{c1}\to \rho^\pm \pi^\mp$ and $\pi\pi K^+K^- >
\pi\pi K^o\bar{K^o}$, which should be
searched for experimentally. The $c\bar{c} n\bar{n}$ component will
lead to an additional contribution to $\eta_c (\pi\pi)_S$ (and $\eta_c
\pi\pi$), and light hadron modes of $\chi_{c1}$. In the former case
this is because $c\bar{c} n\bar{n}$ can decay via OZI allowed diagrams
with one pair creation, while the conventional $c\bar{c}\to
c\bar{c}$ (light hadrons) requires the light hadrons to be created via
two pair creations from two gluons violating the OZI rule. It is known
that $c\bar{c}$ components of $\chi_{cJ}$ cannot describe their
decays, both inclusively and exclusively~\cite{wong}. The light hadron
modes of $\chi_{c1}$ coming from its $c\bar{c}$ component going via
OZI forbidden 
two-gluon annihilation is suppressed by Yang's theorem. A $c\bar{c}
n\bar{n}$ component can have $c\bar{c}$ annihilation into a colour
octet gluon, yielding light hadrons via OZI allowed diagrams. An
additional contribution to measured final states like $2(\pi^+\pi^-)$,
$\pi^+\pi^- K^+ K^-$ and $K^0_S K^+ \pi^-$ is hence expected. It is
noted in passing that threshold mixing with other narrow states should
also be important, e.g. mixing of $\chi_{c0}$, $\chi_{c2}$ and
$\psi´(2S)$ with the $D\bar{D}$ threshold.

 In summary, of the $c\bar{c}$, hybrid and molecular possibilities
considered the $J^{PC}=1^{++}$ assignment for $X$ seems most
promising, because it allows an $S$-wave interaction between the $D^0$
and $D^{\ast 0}$, {\it and} it couples to $J/\psi\: \rho^0$. 
 This resonance can be a 2P resonance shifted by a threshold, 
of genuine molecular
origin, or is generated by a ``shepherd state" scenario~\cite{bcl03}
where the two-meson continuum is driven into a bound state just below
threshold. A $1^{++}$ resonance should be weakly produced in 
$\gamma\gamma$ collisions by Yang's theorem. 

  It is suggested that BES and CLEO-III search for $e^+ e^- \to X$, as
observation will signal $1^{--}$ quantum numbers not expected
here. Also, discovery of $X$ in $p\bar{p} \to X$ at FNAL will indicate
whether $X$ is $J^{PC}$ exotic or not, as $J^{PC}$ exotic quantum
numbers cannot be produced. Central production in e.g. $pp \to pXp$ at
high energy by double Pomeron exchange would confirm $C=+$, since the
Pomeron has $C=+$. The azimuthal angular distribution for production
of $X$ will have a characteristic dependence on $J^P$~\cite{cks}.

This work is supported, in part, by grants from the Particle Physics
and Astronomy Research Council, and the EU-TMR program ``Euridice'',
HPRN-CT-2002-00311 (FEC); and by the U.S. Department of 
Energy contract W-7405-ENG-36 (PRP). We thank Ted Barnes 
and Bob Cahn for discussions.

While this work was in preparation, a discussion of the molecular 
possibility
with similar conclusions to ours has appeared~\cite{torn03}.

\end{document}